\documentclass[aps,prl,reprint,superscriptaddress,longbibliography]{revtex4-2}
\usepackage{blindtext}
\usepackage{graphicx}
\usepackage{amsfonts}
\usepackage{amsmath}
\usepackage{bbold}
\usepackage{physics}
\usepackage{float} 
\usepackage{xcolor}
\usepackage[export]{adjustbox}

\begin{document}

\title{Ballistic Modes as a Source of Anomalous Charge Noise}

\author{Ewan McCulloch}
\affiliation{Department of Electrical and Computer Engineering,
Princeton University, Princeton, NJ 08544, USA}
\affiliation{Department of Physics, University of Massachusetts, Amherst, MA 01003, USA}

\author{Romain Vasseur}
\affiliation{Department of Physics, University of Massachusetts, Amherst, MA 01003, USA}

\author{Sarang Gopalakrishnan}
\affiliation{Department of Electrical and Computer Engineering,
Princeton University, Princeton, NJ 08544, USA}

\begin{abstract}
Steady-state currents generically occur both in systems with continuous translation invariance and in nonequilibrium settings with particle drift. In either case, thermal fluctuations advected by the current act as a source of noise for slower hydrodynamic modes. This noise is unconventional, since it is highly correlated along spacetime rays. We argue that, in quasi-one-dimensional geometries, the correlated noise from ballistic modes generically gives rise to anomalous full counting statistics (FCS) for diffusively spreading charges. We present numerical evidence for anomalous FCS in two settings: (1)~a two-component continuum fluid, and (2)~the totally asymmetric exclusion process (TASEP) initialized in a nonequilibrium state.
%
%
%Ballistic modes are common place in physics, whether as a result of a conserved current as in Galilean invariant or relativistic systems, or otherwise -- for example, as a result of an asymmetry in the microscopic dynamics as occurs in asymmetric simple exclusion process (ASEP) models of traffic flow. If sub-ballistic modes are also present, the ballistic modes will act as a source of ballistically correlated velocity kicks for the slower sub-ballistic mode. %in a convective mechanism similar to that seen in integrable systems. 
%We show that in a Galilean fluid -- a two-component hard-sphere gas -- and in ASEP models, that the full counting statistics (FCS) of a charge mode is anomalous under certain conditions: (1) a quasi-1D geometry; and (2) in the case of multi-lane ASEP, a non-equilibrium initial state such as a step configuration. Using numerical simulations we observe that the charge transfer cumulants strongly violates central limit scaling laws, we observe $C^{\text{step}}_n(t)\sim t^{n/2}$ for a step initial and $C^{\text{eq}}_n(t)\sim t^{n/4}$ in equilibrium for hard spheres.
%
\end{abstract}

\maketitle

Chaotic systems generically thermalize~\cite{nandkishore2015many}. When their dynamics is local, thermalization occurs locally: the system rapidly approaches an equilibrium state characterized by smoothly varying densities of its conserved charges, which then relax by hydrodynamic mechanisms. Conventionally, hydrodynamics was developed (and tested) as a framework for computing expectation values. Fluctuating hydrodynamics extends this framework to predict the full spatial distribution of charge densities~\cite{spohn2012large, liu2018lectures}, which can be characterized through the full counting statistics (FCS)~\cite{Levitov1993, Levitov1996ElectronCS, de_Jong_1996, Nagaev2_2002,  Levitov_2004, Belzig2002, Belzig2001, Nazarov_2002, RevModPhys.81.1665, Sch_nhammer_2007, hofferberth2008probing, kitagawa2011dynamics, wei2022quantum, rosenberg2023dynamics, wienand2023emergence, 2022arXiv220411680H, PhysRevB.87.184303, Najafi2017, RevModPhys.81.1665, Groha2018, 2023arXiv231202929H}, i.e., the probability distribution of the total charge flux across a given surface in time $t$. The structure of large-scale fluctuations in fluctuating hydrodynamics is called macroscopic fluctuation theory (MFT)~\cite{Bertini_2015, Lazarescu_2015, Bernard_2021, Mallick_2022}, and was recently extended to ballistic systems~\cite{BallisticMFT, Myers2018}. The predictions of fluctuating hydrodynamics and MFT go beyond the thermalization of local expectation values, since they involve nonlocal many-point correlation functions; they are, however, supported by numerics, arguments based on random circuits~\cite{PhysRevLett.131.210402}, and recent experimental studies in ultracold gases~\cite{wienand2023emergence}. 

Even within fluctuating hydrodynamics, identifying the late-time behavior of FCS remains a largely open question, outside of a few exactly solvable cases such as the one-dimensional symmetric~\cite{Derrida_2009, Mallick_2022, Bernard2022} and totally asymmetric~\cite{PhysRevLett.104.230602} simple exclusion processes (labeled SSEP and TASEP respectively). SSEP is a paradigm for standard diffusive transport, and its FCS is what one might expect for diffusive systems: in particular, all cumulants of the charge flowing across a point over a time $t$ scale with the same exponent $t^{1/2}$, so typical fluctuations are asymptotically Gaussian. To leading order this behavior is consistent with a naive picture in which one takes the system to have thermalized over a region of size $\sim t^{1/2}$ on a timescale $t$, and the FCS is just the thermodynamic fluctuations of the charge in that region. Recently, it was realized that this behavior is \emph{not} in fact universal for systems with diffusive transport: in solvable cellular automata~\cite{Krajnik2022a}, integrable XXZ spin chains~\cite{Krajnik2022b, Gopalakrishnan2022a}, and stochastic models of single-file diffusion~\cite{Krajnik2022c}, a very different limiting behavior for the FCS was observed, with persistently nongaussian fluctuations. 

While the initially studied models were integrable, we recently argued~\cite{gopalakrishnan2024non} that this anomalous FCS has a purely hydrodynamic origin, see also Ref.~\cite{2024arXiv240620091Y}. We focused on the case of Dirac fluids, where the anomalous FCS comes from the interplay between Lorentz invariance and particle-hole symmetry. In the present work, we extend this analysis to much more general settings, which---unlike the Dirac fluid---allow us to test our predictions against efficient classical numerics. The minimal ingredients for our analysis are a system with two conserved charges, of which the first moves ballistically or is in a current-carrying nonequilibrium steady state, while the second diffuses. Our main result is that the hydrodynamic coupling between these two charges endows the second charge with anomalous FCS. The cumulants of charge transferred across a cut scale as $C^{\text{step}}_n(t)\sim t^{n/2}$ for an initial state with a step density profile and $C^{\text{eq}}_n(t)\sim t^{n/4}$ for an equilibrium initial state. We support these conclusions with numerical studies of three models: (1)~a two-component classical fluid of hard spheres, (2)~a variant of TASEP in which the particles have two color labels, and (3)~a model of two coupled chains, each governed by TASEP with a distinct asymmetry parameter, which causes relative drift between the particles in the chains.

The origin of anomalous FCS in these models can be understood as follows. The ballistic (or current-carrying) modes have thermal fluctuations, which interact with the diffusive modes. Since the two modes move at parametrically different velocities, the diffusive modes in a given region only encounter each ballistic fluctuation once, so for the purposes of linear-response transport the ballistic fluctuations act as Markovian noise, and just renormalize the diffusion constant. However, this noise is ballistically propagating and is therefore perfectly correlated along spacetime rays. These correlations mean that (in contrast with a standard diffusive system like SSEP) diffusive modes in different parts of the system experience the same fluctuations translated along light rays, giving rise to a parametric enhancement of noise. (We note that parametric noise enhancement can also occur near critical points \cite{2014PhRvB..90r0505B}, or as a result of long-range, rather than ballistic, correlations at low temperatures \cite{2006ForPh..54..917B}.)

\emph{Two-component classical fluids}.---We first consider a two-component fluid, with ``charge'' components $+$ and $-$, confined in a quasi-one-dimensional tube~\footnote{Note that the ``charge'' is fictitious, there are \emph{no} Coulomb interactions.}. Each species is separately conserved. The particles in the fluid collide with one another elastically, so the evolution of the fluid is chaotic but deterministic. Moreover, the equations of motion obey continuous translation invariance along the tube. We sample initial states of the fluid at random from a thermal distribution, and compute the FCS of the ``charge''---i.e., the total flux of $+$ particles minus the total flux of $-$ particles---across a fixed cross section of the tube. 
Nontrivial FCS can arise only in quasi-one-dimensional geometries: in higher dimensions, the charge transfer across a surface occurs independently along each part of the surface, so the statistics of total charge transfer is set by the central limit theorem regardless of the real local mechanism. 
While this forces us to consider quasi-one-dimensional geometries, we choose a tube that is wide enough so that the kinematics of collisions is higher-dimensional: i.e., the distribution of particle momenta equilibrates. 

The two-component classical fluid has the following conserved densities: energy $\varepsilon$, momentum $\boldsymbol{\phi}$, the total particle number $n \equiv n_+ + n_-$, the imbalance $q \equiv n_+ - n_-$. We will also impose particle-hole symmetry (PHS): both the equilibrium measure and the interparticle interactions are invariant under swapping the charge of every particle. Under these conditions we can write down the following set of continuity equations,
\begin{align}
\partial_t n +\nabla\cdot \boldsymbol{\phi} &= 0,
\quad \partial_t \phi_i + \partial_j g_{ij} = 0, \nonumber\\
\partial_t q + \nabla\cdot \boldsymbol{j}_q &= 0,\quad 
\partial_t \varepsilon + \nabla \cdot \boldsymbol{j}_\varepsilon = 0,
\end{align}
where $\boldsymbol{j}_\varepsilon$ is the energy current and $g_{ij}$ is the current of $\boldsymbol{\phi}$. 

We now write down the corresponding constitutive relations. We begin with the Euler-scale relations (i.e., those with no spatial derivatives). Since $g_{ij}$ is a rank $2$ tensor which is even under spatial inversion and time reversal, the only terms at linear order in the constitutive relation are $g_{ij} = (a n(\boldsymbol{x},t) + b \varepsilon(\boldsymbol{x},t)) \delta_{ij}$. Similarly, the energy current is a rank $1$ tensor that is odd under spatial and time inversion. Therefore, at the linear order the energy current only overlaps with particle current, i.e., $\boldsymbol{j}_\varepsilon = c \boldsymbol{\phi}(\boldsymbol{x},t)$. To specialize to quasi-one-dimensional geometries, we recall that only the component of $\boldsymbol{\phi}$ along the tube is conserved. To summarize, the system of equations for $n, \varepsilon, \boldsymbol{\phi}$ is precisely what one would have for a one-component fluid. In one dimension, these equations give rise to two sound modes with finite velocity and a heat mode with zero velocity. Once nonlinearities, diffusion, and noise are included, the sound mode broadens as $t^{2/3}$ and the heat mode broadens as $t^{3/5}$~\cite{PhysRevLett.56.889, PhysRevLett.108.180601, SpohnNLFH,PhysRevLett.54.2026, PhysRevA.92.043612,Fibonacci}. 

The charge current has the same symmetry requirements as the particle and energy density in addition to the PHS which, crucially, precludes any overlap with the particle current. The leading order contribution to the charge current is the quadratic term $\boldsymbol{j}_q \sim  q(\boldsymbol{x},t) \boldsymbol{\phi}(\boldsymbol{x},t)$ (we have not included a term $q\boldsymbol{j}_\varepsilon$ as at the Euler scale the energy current is already proportional to $\boldsymbol{\phi}$). In addition to this quadratic term, one can write down conventional Fickian diffusion and noise terms, $-D \partial_x q + \xi$. The lowest-order term through which charge fluctuations can couple back to the energy (or particle-number) fluctuations is  $\varepsilon \sim q^2$. This term is subleading to the standard nonlinearities that broaden the sound and heat modes~\cite{Fibonacci}, and we will not consider it further. To summarize, $q$ obeys the hydrodynamic equation
\begin{equation}\label{redux}
\partial_t q + \partial_x (C_1 \phi q - D \partial_x q + \xi) = 0,
\end{equation}
up to higher order terms. Here the noise term $\xi$ is white noise with a strength related to $D$ by the fluctuation-dissipation theorem. The structure of these equations is exactly the same as that studied in Ref.~\cite{gopalakrishnan2024non}, except for the presence of the heat mode in the present case. However, since $q$ directly couples only to $\phi$, which is orthogonal to the heat mode, the coupling between $q$ and the heat mode first enters at cubic order, as $j_q \sim q\phi + \order{q\varepsilon\phi}$, which only contributes logarithmic corrections \cite{Devillard1992} which we ignore.

\begin{figure}[!t]
    \centering
    \includegraphics[width = 0.48\textwidth,trim=0cm 20cm 10cm 0cm,]{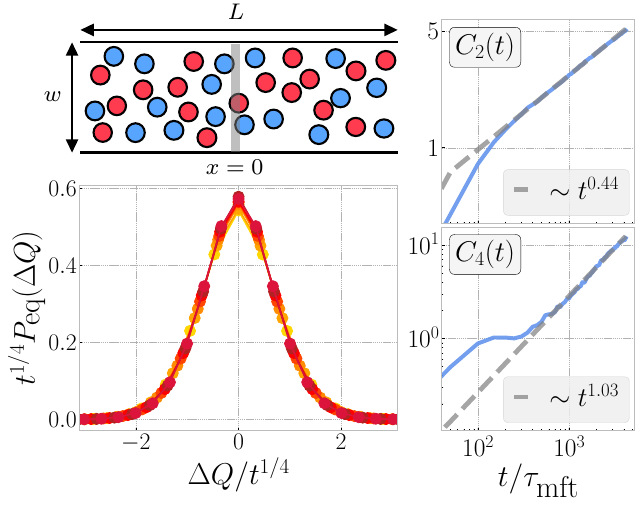}
    \caption{\textbf{Hard disks:} (top left) a cartoon of a gas of hard disks confined on a strip of width $w=6.15d$ ($d$ is the disk diameter). (bottom left) the equilibrium charge transfer probability distribution $P_t(\Delta Q)$ at various times. (top right) the second and fourth (bottom right) charge transfer cumulants in equilibrium. Times are given in units of the mean-free-time $\tau_{\textrm{mft}}\approx 0.0329\textrm{s}$.}
    \label{fig:HardDisksEqm}
\end{figure}

The key observation in Ref.~\cite{gopalakrishnan2024non} is that the fluctuations in $\phi$ and $q$ decouple, because $\phi$ is propagating ballistically. Therefore, one can treat $\phi$ as ballistically propagating noise, with a correlation function of the form $\delta(x - vt)$ where $v$ is the speed of sound. Integrating Eq.~\eqref{redux} with ballistically propagating noise, one finds that all the cumulants of the charge transfer scale as $t^{n/2}$ (for a nonequilibrium initial state) and as $t^{n/4}$ (for an equilibrium initial state)~\cite{gopalakrishnan2024non}. %For the special case where the Fick's law term and its associated white noise are zero, one has an explicit expression for the distribution of transferred charge:
%
%[write in general expression] {\color{red} i'd remove this sentence.}

\emph{Numerical results}.---To test these predictions numerically we consider the hard sphere gas \cite{2023arXiv230502452R}, a fluid of spheres of diameter $d$ which experience elastic collisions. Specifically we consider two-component hard disks (two dimensional hard spheres) with both species having the same diameter $d=1$ to retain the PHS. We then evolve the hard-spheres on a thin strip of width $w=6.15$ and length $L=20000$ (see Fig.~\ref{fig:HardDisksEqm}) using event-driven molecular dynamics simulations \cite{2010arXiv1004.3501B}
while sampling from an equilibrium steady state with temperature $k_{B}T = 1.5$ and packing density $\rho=0.75$. We measure the transfer of charge $\Delta Q$ across a cut at the center of the strip at different times $t$ for $N=3650$ samples to form the probability distribution $P_t(\Delta Q)$.

The distribution $P_t(\Delta Q)$ shows a clear scaling collapse with the scaling $\Delta Q \sim t^{1/4}$ (see Fig.~\ref{fig:HardDisksEqm}). In Fig.~\ref{fig:HardDisksEqm}, we show the second and fourth cumulants, which are observed to have the scaling $C_2(t)\sim t^{0.44}$ and $C_4(t)\sim t^{1.03}$. This is in good agreement with the predictions for a convective diffusion mechanism $C_n(t)\sim t^{n/4}$.

\emph{TASEP with charge labels}.---So far, we have considered anomalous FCS in systems that have a ballistic mode either because of integrability or by symmetry. A natural question is whether any symmetry beyond PHS is \emph{necessary} for seeing anomalous FCS. To demonstrate that it is not, we turn to models with no symmetries, in which the persistent flow of one species is instead a consequence of the system being out of equilibrium. First, we will study charge FCS 
%\textbf{ASEP with charge labels} -- 
%In the previous section, we studied the charge transfer FCS for a diffusive charge mode in a Galilean invariant system. In this section we will drop this symmetry and study charge FCS 
in a model with the minimal ingredients for convective diffusion, i.e., a charge mode and a single ballistic mode.

We consider a single-lane TASEP where the particles carry a charge $q=\pm 1$. The labels are only spectators to the usual TASEP dynamics, which has particles hop to an empty site to the right with a rate $r$ (which we will take to be $r=1/2$) and no hopping otherwise. The charge mode has a PHS at half filling, ensuring that the charge current is given at leading order by the same quadratic coupling that we saw in the classical gas. 

The steady state particle current for this TASEP is given by $j=r n(1-n)$ \cite{LiggettBook,Domb:1990404} where $n$ is the total particle density. The charge mode is carried along with the overall particle current at a velocity $v_q = j/n = r(1-n)$, while the usual TASEP mode (the total particle density mode) travels at a generically different velocity given by $v_n = \partial j/\partial n = r(1-2n)$. By moving into the fluid frame, we can engineer a system in which a sub-ballistic charge mode $q(x,t)$ coexists with a ballistic mode $n(x,t)$.

In the fluid frame, the charge current at the Euler scale is simply given by $j_q = d q(x,t) n(x,t) + \cdots$ where ellipsis represents cubic order terms and higher. %This is precisely the same non-linearity we have seen in the two-component fluid. Moreover,
The inability for the charge labels to reorder means that the diffusion constant in Fick's law term $j_q = -D\partial_x q(x,t)$ is zero. In this case, treating the ballistic mode $n(x,t)$ as ballistically correlated noise, the charge transfer distribution is known exactly (within a diffusive scaling limit)~\cite{gopalakrishnan2024non},
\begin{equation}\label{eqm_dist_theory}
P_t(\Delta Q) = \int_0^\infty dX \frac{{\rm exp} \left( - \frac{X^2}{4 \alpha t} - \frac{\Delta Q^2}{2 X} \right)}{\sqrt{2 \pi^2  } \sqrt{\alpha t X}},
\end{equation}
where $\alpha$ is a parameter determined by thermal fluctuations in the initial state. 

\begin{figure}[!t]
    \centering
    \includegraphics[width = 0.45\textwidth,trim=0cm 1cm 0cm 0cm,]{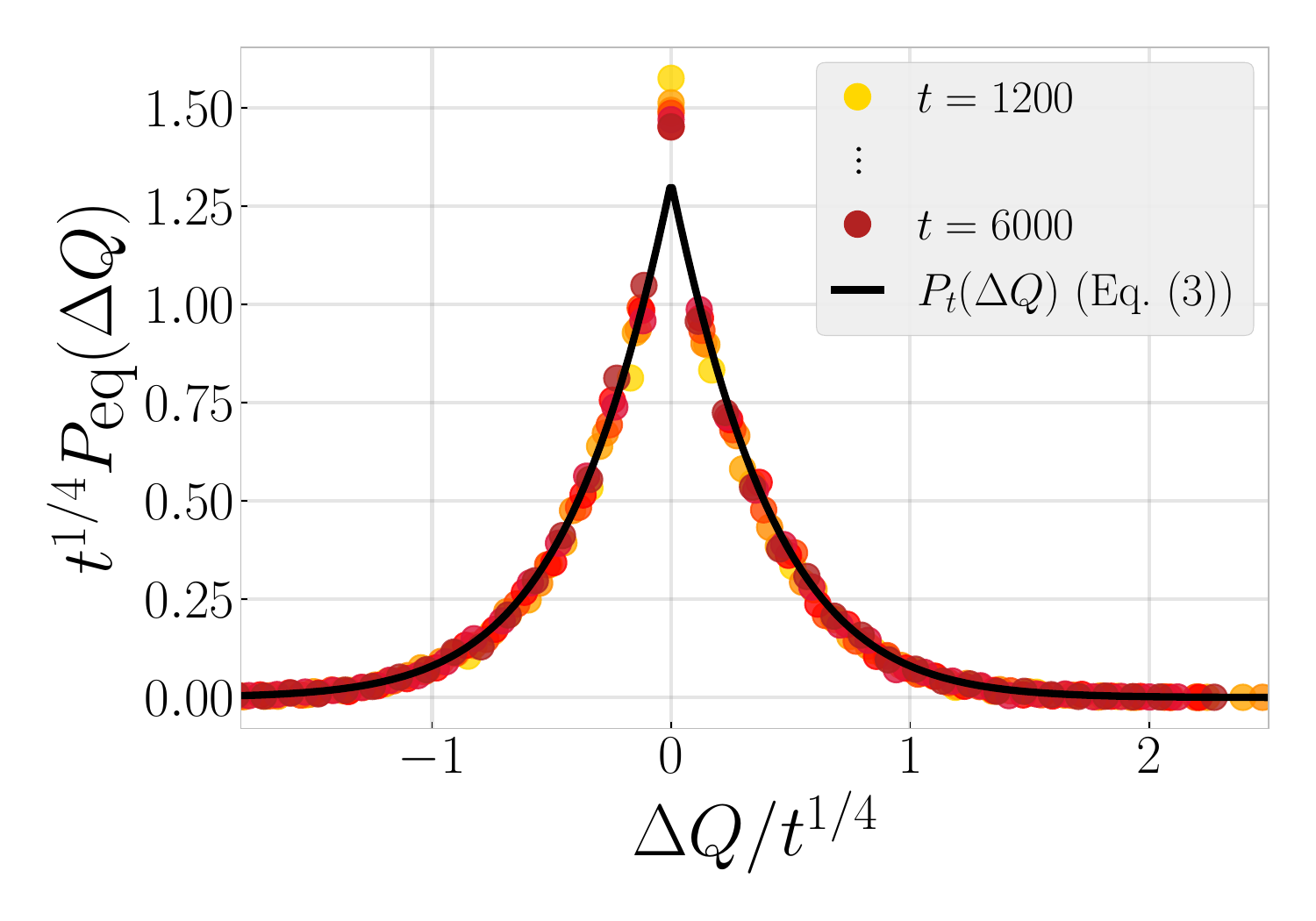}
    \caption{\textbf{Charged TASEP in equilibrium:} The charge transfer statistics for a single-lane TASEP in equilibrium where the particles can carry a charge $q=\pm1$. The charge transfer is measured in a frame moving with the particle current. The black curve is the theoretical prediction for $P_t(\Delta Q)$ (Eq.~\ref{eqm_dist_theory}) with $\alpha=0.036$ determined through a fit to the data.}
    \label{fig:TASEP_eqm}
\end{figure}

\begin{figure}[!t]
    \centering
    \includegraphics[width = 0.45\textwidth,trim=0cm 1cm 0cm 0cm,]{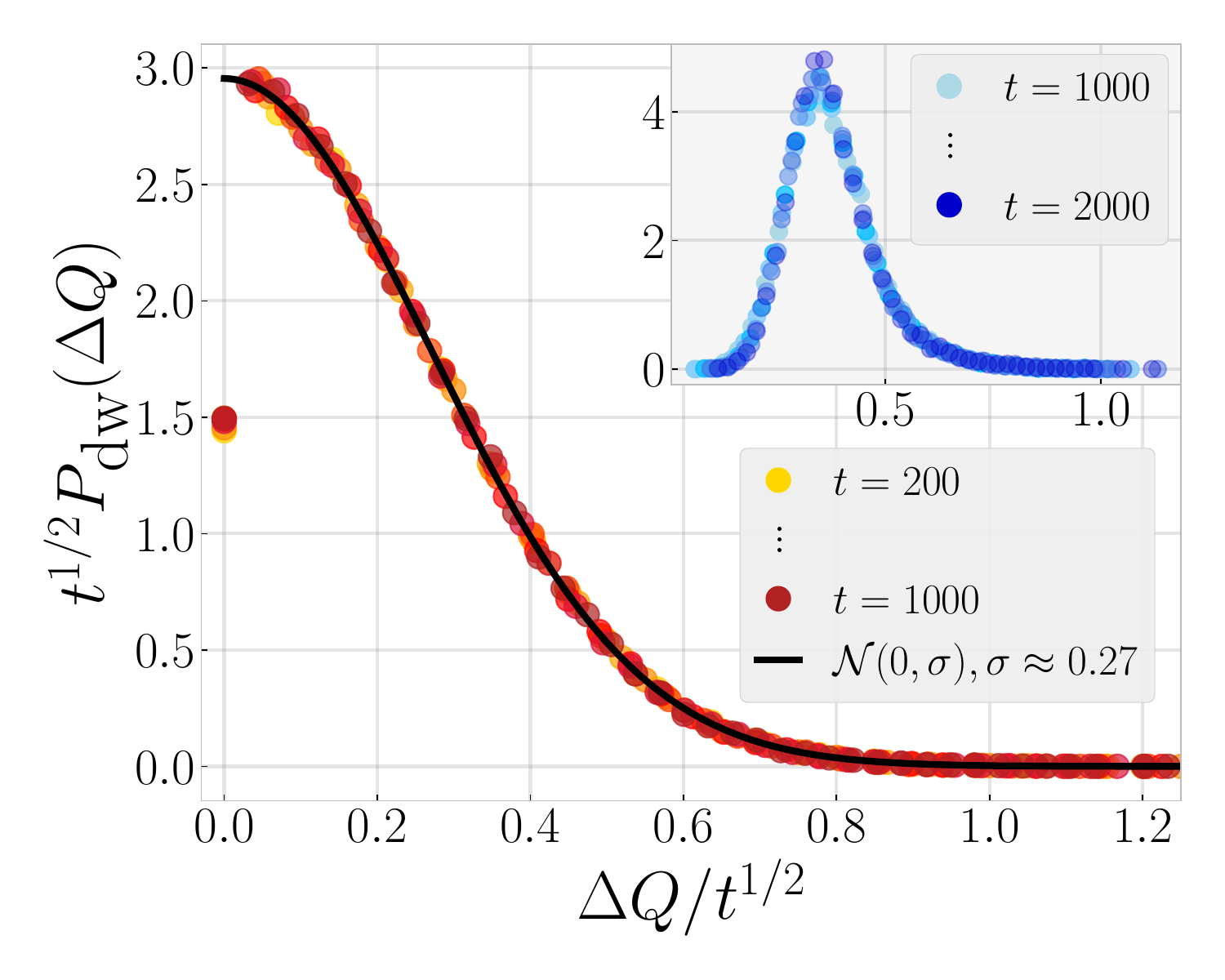}
    \caption{\textbf{Charged TASEP with a domain-wall initial:} (Main) The charge transfer statistics for a single-lane TASEP with a domain-wall initial for the charge labels $q=\pm 1$. As in equilibrium, the charge transfer is measured in a frame moving with the particle current. The $t\to\infty$ asymptotic distribution is predicted to be a half Gaussian~\cite{Gopalakrishnan2022a}. (Inset) By allowing particles to exchange charge labels when they collide, the domain wall is softened over time, and the non-analyticity at $\Delta Q = 0$ is removed.}
    \label{fig:TASEP_dw}
\end{figure}

We simulate the single-lane TASEP with particle density $n=0.6$ and system size $L=2000$ for times up to $t=6000$. The equilibrium charge transfer distribution shows a scaling collapse with $\Delta Q \sim t^{1/4}$ as shown in Fig.~\ref{fig:TASEP_eqm} and shows good agreement with the prediction in Eq.~\ref{eqm_dist_theory}. For a domain-wall initial state in the charge labels, the same particle density $n=0.6$ and a system size $L=1000$, we find a scaling collapse that agrees very well with a half-Gaussian, as predicted for convective diffusion in absence of a Fick's law diffusion term~\cite{gopalakrishnan2024non}. This is shown in Fig.~\ref{fig:TASEP_dw}.

By allowing particles to exchange their charge labels with a probability $p$ at each collision, the Fick's law diffusion constant $D$ can be restored and the limit shape for the charge transfer distribution changes, with the non-analyticity at $\Delta Q=0$ softened. We show this for $p=0.15$ in the inset of Fig.~\ref{fig:TASEP_dw}.

\emph{Two-lane TASEP}.---The final setting we consider is another non-Galilean system, that of a two-lane TASEP. In standard TASEP, the particles are not charged and the system has only a single mode. In order to have a ballistic mode for our chosen `charge mode' (which could be any conserved density) to couple to, one can just couple multiple TASEP lanes together. 
However, this setting differs from the two component classical gas and the charged single lane TASEP in a crucial way: it has no particle-hole symmetry. This means that the particle currents $j_{i}$ (for each lane $i$) will generically have the KPZ non-linearities $j_i\sim n_i^2$. If one takes the particle number for one of the lanes as the charge, the KPZ fluctuations in the equilibrium particle transfer, $\Delta N \sim t^{1/3}$, would overwhelm the contribution from convective diffusion, which scales with a smaller power, $\Delta N \sim t^{1/4}$.

In order to make the convective term the dominant contribution, we prepare a given lane in a domain-wall configuration with different densities $n_L$ and $n_R$ in the left and right of the system. The convective term then shifts the entire particle configuration by some amount $\sqrt{t}$ over a time $t$, giving particle transfer fluctuations $\sim t^{1/2}$. The domain wall is smeared over some distance $\sim t^{2/3}$ due to the KPZ broadening and the corresponding fluctuations in the particle transfer scale as $\sim t^{1/3}$, which are sub-leading to the convective term. We therefore expect to find the anomalous scaling $C_n(t)\sim t^{n/2}$ in the particle transfer cumulants (apart from $C_1(t)\sim t$ which encodes the overall particle current).

\begin{figure}[t]
    \centering
    \includegraphics[width = 0.45\textwidth,trim=0cm 2.5cm 0cm 0cm,]{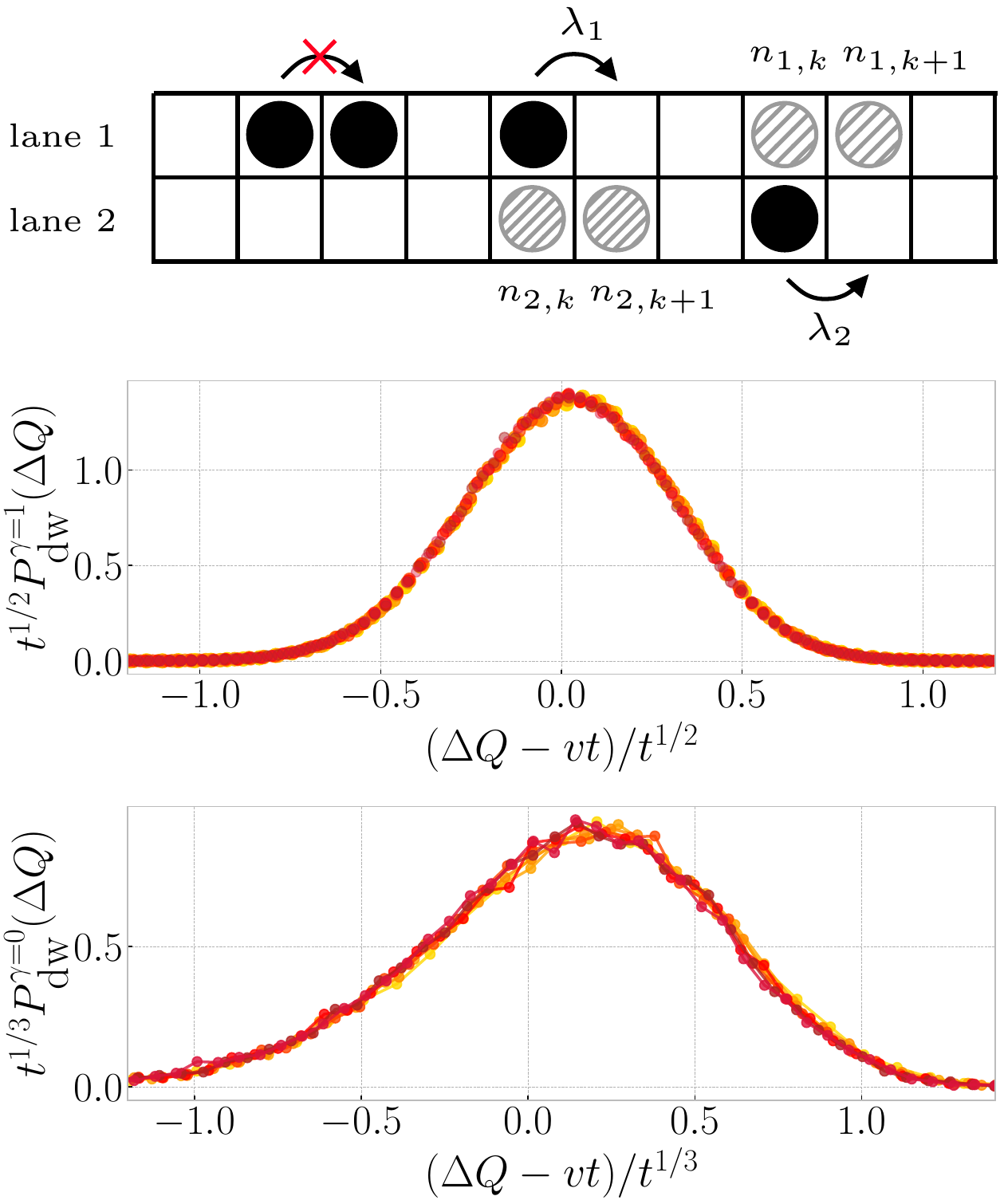}
    \caption{\textbf{Two-lane TASEP:} (top) A two-lane totally asymmetric exclusion process with occupancy dependent hopping rates $\lambda_1$ and $\lambda_2$. The striped circles indicate that the hopping rates depend on whether these sites are vacant or occupied. (middle) Coupled TASEP with domain-wall initial: particle transfer distribution $P^{\gamma=1}_{\textrm{dw}}(\Delta Q(t))$ for lane $1$. Lane $1$ is initialized with a step density distribution and lane $2$ initialized with a flat density distribution. The color indicate times -- yellow for earlier times ($t=500$) and red for later times ($t=1000$). Uncoupled TASEP with domain-wall initial: the particle transfer distribution $P^{\gamma=0}_{\textrm{dw}}(\Delta Q(t))$ for a single-lane TASEP. The system is initialized with a step density distribution. The color indicate times -- yellow for earlier times ($t=1000$) and red for later times ($t=4000$). }  
        \label{fig:asepladder}
\end{figure}

We consider the same stochastic lattice gas as introduced in \cite{2014PhRvL.112t0602P}. This model has particles hopping on a two-lane ladder with no hops between the lanes so that the particle number on each lane is separately conserved and the hydrodynamic description has two modes. The particles have totally asymmetric hard-core exclusion interactions, i.e., each site can host at most one particle, and the hops occur only in the positive direction (Fig.~\ref{fig:asepladder}). %The allowed hoppings are summarized in Fig.~\ref{fig:hopping}.
The hopping rates for the two lanes are tunable by a facilitated hopping parameter $\gamma$ and a lane bias $b$, and for position $k$ are given by $\lambda_{1,k} = 1 + \frac{\gamma}{2} (n_{2,k}+n_{2,k+1})$ and $\lambda_{2,k} = b + \frac{\gamma}{2} (n_{1,k}+n_{1,k+1})$ for lanes $1$ and $2$ respectively.
%\begin{equation}\label{hopping}
%    \lambda_1 = 1 + \frac{\gamma}{2} n_1, \quad \lambda_2 = b + \frac{\gamma}{2} n_2.
%\end{equation}
For $\gamma \geq -\min(1,b)$ (the parameter regime we work in) this model is a TASEP.

%\begin{figure}[!t]
%    \centering
%    \includegraphics[width = 0.43\textwidth]{hopping.png}
%    \caption{A two-lane totally asymmetric exclusion process with occupancy dependent hopping rates $\lambda_1$ and $\lambda_2$. The striped circles indicate that the hopping rates depend on whether these sites are vacant or occupied.}
%    \label{fig:hopping}
%\end{figure}

We perform Monte Carlo simulations for coupled TASEP of length $L=2000$ with parameters $b=1/2$, $\gamma=1$, and uncoupled TASEP with $b=1/2$, $\gamma=0$. In both cases, we take the initial state to be drawn from a domain-wall density profile in the first lane with $n_L=1/2$ and $n_R=0$ and a flat density profile for the second lane with $n_2=0.6$. In the coupled case, ballistic waves travel along the system and scatter off of the domain wall which becomes broadened and receives random displacements. %The random displacements from incoming ballistic modes is another example of convective mechanism we have already seen in the hard disk gas and in the single lane ASEP with charges. 
In the uncoupled case, the domain wall also broadens due to the KPZ non-linearity, but does not receive any random displacements.

Particle transfer is measured in the first lane across a cut stationary in the lab frame. The subsequent distributions show a scaling collapse with $(\Delta Q -vt)\sim t^{1/2}$ for the coupled TASEPs and a collapse with $(\Delta Q -ut)\sim t^{1/3}$ for the uncoupled case as shown in Fig.~\ref{fig:asepladder}, where $v$ and $u$ are the characteristic propagation velocities for the domain wall in the coupled and uncoupled case respectively, and which are easily extracted from the mean charge transfer. This result clearly demonstrates that the fluctuations from the convective mechanism dominates over KPZ fluctuations for a non-equilibrium initial state.

\emph{Discussion}.---In this work we demonstrated, using hydrodynamic arguments and numerical simulations, that anomalous, non-gaussian FCS is a generic consequence of coupling slow hydrodynamic modes to ballistically propagating fluctuations. 
We verified that anomalous FCS occurs in three increasingly unstructured models: first, closed chaotic systems with Galilean invariance and particle-hole symmetry; second, open nonreciprocal systems with particle-hole symmetry, in which the ballistic mode arises because the dynamics breaks inversion symmetry; and third, open systems without particle-hole symmetry, initialized in domain-wall initial states. 
These results establish the generality of the convective mechanism for anomalous FCS, under conditions that are far weaker than those stipulated in Ref.~\cite{gopalakrishnan2024non}. 

Since we focused on FCS, our results are limited to quasi-one-dimensional geometries. However, ballistically correlated noise occurs in any dimensionality. Identifying diagnostics---such as angle-dependent noise correlations---that can detect anomalous fluctuations in these geometries is an interesting task for future work. A closely related task would be to interpret our hydrodynamic arguments in field-theoretic terms, i.e., to characterize the fixed-point theories with coexisting ballistic and diffusive modes that give rise to anomalous FCS~\cite{2023arXiv230403236D}.

\emph{Acknowledgments}.--- We thank Luca Delacr\'etaz,  Nathalie de Leon, Jacopo De Nardis, Matthew Foster, Paolo Glorioso, David Huse, Vedika Khemani, Andrew Lucas, Alan Morningstar, Rahul Sahay, and Frank Zhang for helpful discussions, collaboration on previous works  and/or comments on this manuscript.
We acknowledge support from NSF Grants No. DMR-2103938 (S.G., E.M.) and DMR-2104141 (E.M., R.V.).

\bibliography{Refs}

\end{document}